\documentclass[aps,pre,amsmath,lengthcheck,superscriptaddress]{revtex4-2}

\usepackage{graphicx}\graphicspath{ {figures/} }
\usepackage{hyperref}
\hypersetup{colorlinks,allcolors=blue,breaklinks}

\newcommand{\be}{\begin{equation}}
\newcommand{\ee}{\end{equation}}
\newcommand{\ba}{\begin{align}}
\newcommand{\ea}{\end{align}}
\newcommand{\bi}{\begin{itemize}}
\newcommand{\ei}{\end{itemize}}

\newcommand{\la}{\left\langle}
\newcommand{\ra}{\right\rangle}
\newcommand{\pd}{\partial}

\newcommand{\bla}{bla\\bla\\bla\\bla\\bla}

\newcommand{\mb}[1]{\mbox{\boldmath$#1$}}
\newcommand{\mc}[1]{\mathcal{#1}}

\begin{document}

\title{Optimal work fluctuations for finite-time and weak processes}

\author{Pierre Naz\'e}
\email{pierre.naze@unesp.br}

\affiliation{\it Universidade Estadual Paulista, 14800-090, Araraquara, S\~ao Paulo, Brazil}

\date{\today}

\begin{abstract}

The optimal protocols for the irreversible work achieve their maximum usefulness if their work fluctuations are the smallest ones. In this work, for classical and isothermal processes subjected to finite-time and weak drivings, I show that the optimal protocol for the irreversible work is the same for the variance of work. This conclusion is based on the fluctuation-dissipation relation $\overline{W}=\Delta F+\beta \sigma_W^2/2$, extended now to finite-time and weak drivings. To illustrate it, I analyze a white noise overdamped Brownian motion subjected to an anharmonic stiffening trap for fast processes. By contrast with the already known results in the literature for classical systems, the linear-response theory approach of the work probabilistic distribution is not a Gaussian reduction.

\end{abstract}

\maketitle

\section{Introduction}
\label{sec:intro}

The optimization of the thermodynamic work of driving processes is a practical example where averages and fluctuations work side-by-side. Finding a protocol to which the external parameter of the system leads to the minimal value of the average of the thermodynamic work has its utmost value when its fluctuations are equally minimal. 

Few remarkable works have been done in the last decades unveiling relations between those concepts involving averages and fluctuations of the thermodynamic work. In the classical realm, Jarzynski \cite{jarzynski1997nonequilibrium} found out from its famous equality that systems with time-dependent quadratic potentials, starting its driving process in contact with a heat bath of temperature $\beta^{-1}$, obey the following fluctuation-dissipation relation
\be
\overline{W}=\Delta F+\frac{\beta}{2}\sigma_W^2,
\label{eq:wfdt}
\ee
where $\overline{W}$ is the average work, $\sigma_W^2$ is the variance of the work, and $\Delta F$ is the difference of Helmholtz free energy between the final and initial equilibrium states of the process. Some years later, Speck and Seifert \cite{speck2004distribution} deduced the same result, but now for slowly-varying processes, that is, processes whose rate is not fast enough when compared to the relaxation rate of the system. Very recently, in the quantum realm, Miller and coauthors found out that such fluctuation-dissipation relation fails when the coherence of quantum systems is added \cite{miller2019work}. From the point-of-view of optimization, in these regimes where the fluctuation-dissipation relation holds, the minimal work is the precisest one. 

The objective of this work is to derive the same result for isothermal, finite-time, and weak driving processes, where the rate of the process is arbitrary and the perturbation of the external parameter is small compared to its initial value. To accomplish that, I generalize the fluctuation-dissipation relation \eqref{eq:wfdt} to this regime using linear-response theory. To illustrate such a relation, I analyze the case of the overdamped Brownian motion subjected to an anharmonic stiffening trap and white noise for fast processes. Finally, as a main consequence of our findings, the Euler-Lagrange equation that determines the optimal protocol for the minimization of the variance of the work is the same as the one that minimizes the irreversible work.

\section{Preliminaries} 
\label{sec:preliminaries}

I start defining notations and developing the main concepts to be used in this work. This section is based on the technical introductory section of Ref.~\cite{naze2022optimal}.

Consider a classical system with a Hamiltonian $\mc{H}(\mb{z}(\mb{z_0},t)),\lambda(t))$, where $\mb{z}(\mb{z_0},t)$ is a point in the phase space $\Gamma$ evolved from the initial point $\mb{z_0}$ until time $t$, with $\lambda(t)$ being a time-dependent external parameter. During a switching time $\tau$, the external parameter is changed from $\lambda_0$ to $\lambda_0+\delta\lambda$, with the system being in contact with a heat bath of temperature $\beta\equiv {(k_B T)}^{-1}$, where $k_B$ is Boltzmann's constant. The average work performed on the system during this interval of time is
\be
\overline{W} \equiv \int_0^\tau \la\overline{\pd_{\lambda}\mc{H}}(t)\ra_0\dot{\lambda}(t)dt,
\label{eq:work}
\ee
where $\partial_\lambda$ is the partial derivative in respect to $\lambda$ and the superscripted dot the total time derivative. The generalized force $\la\overline{\pd_{\lambda}\mc{H}}\ra_0$ is calculated using the averaging $\overline{\cdot}$ over the stochastic path and the averaging $\langle\cdot\rangle_0$ over the initial canonical ensemble. The external parameter can be expressed as
\be
\lambda(t) = \lambda_0+g(t)\delta\lambda,
\label{eq:ExternalParameter}
\ee
where to satisfy the initial conditions of the external parameter the protocol $g(t)$ must satisfy the following boundary conditions
\be
g(0)=0,\quad g(\tau)=1. 
\label{eq:bc}
\ee

Linear-response theory aims to express average quantities until the first-order of some perturbation parameter considering how this perturbation affects the observable to be averaged and the probabilistic distribution \cite{kubo2012}. In our case, we consider that the parameter does not considerably change during the process, $|g(t)\delta\lambda/\lambda_0|\ll 1$, for all $t\in[0,\tau]$ and $\lambda_0\neq 0$. The generalized force can be approximated until the first order as
\begin{equation}
\begin{split}
\la\overline{\pd_{\lambda}\mc{H}}(t)\ra_0 =&\, \la\pd_{\lambda}\mc{H}\ra_0-\delta\lambda\widetilde{\Psi}_0 g(t)\\
&+\delta\lambda\int_0^t \Psi_0(t-t')\dot{g}(t')dt',
\label{eq:genforce-relax}
\end{split}
\end{equation}
where 
\be
\Psi_0(t) = \beta\la\pd_\lambda\mc{H}(0)\overline{\pd_\lambda\mc{H}}(t)\ra_0-\mc{C}
\ee 
is the relaxation function and $\widetilde{\Psi}_0\equiv \Psi_0(0)-\la\pd_{\lambda\lambda}^2\mc{H}\ra_0$ \cite{kubo2012}. The constant $\mc{C}$ is calculated to vanish the relaxation function for long times \cite{kubo2012}. Combining Eqs.~\eqref{eq:work} and \eqref{eq:genforce-relax}, the average work performed at the linear response of the generalized force is
\begin{equation}
\begin{split}
\overline{W} = &\, \delta\lambda\la\pd_{\lambda}\mc{H}\ra_0-\frac{\delta\lambda^2}{2}\widetilde{\Psi}_0\\
&+\delta\lambda^2 \int_0^\tau\int_0^t \Psi_0(t-t')\dot{g}(t')\dot{g}(t)dt'dt.
\label{eq:work2}
\end{split}
\end{equation}

We observe that the double integral on Eq.~\eqref{eq:work2} vanishes for long switching times \cite{naze2020}, which indicates that the other terms are the contribution of the difference of Helmholtz's free energy. The irreversible work $W_{\rm irr}$ is therefore
\begin{equation}
\begin{split}
W_{\rm irr} = \frac{\delta\lambda^2}{2} \int_0^\tau\int_0^\tau \Psi_0(t-t')\dot{g}(t')\dot{g}(t)dt'dt,
\label{eq:wirrLR}
\end{split}
\end{equation}
where the symmetric property of the relaxation function was used \cite{kubo2012}. The regime where such expression holds is the finite-time and weak processes, where the ratio $\delta\lambda/\lambda_0\ll 1$, while $\tau_R/\tau$ is arbitrary.

\section{Optimization problem}

Consider the irreversible work rewritten in terms of the protocols $g(t)$ instead of its derivative
\begin{align}
    W_{\rm irr} =& \frac{\delta\lambda^2}{2}\Psi_0(0)+\delta\lambda^2\int_0^\tau \dot{\Psi}_0(\tau-t)g(t)dt\\&-\frac{\delta\lambda^2}{2}\int_0^\tau\int_0^\tau \ddot{\Psi}_0(t-t')g(t)g(t')dt dt'.
\end{align}
Using the calculus of variations, one can derive the Euler-Lagrange equation that furnishes the optimal protocol $g^*(t)$ that minimizes the irreversible work \cite{naze2022optimal}
\be
\int_0^\tau \ddot{\Psi}_0(t-t')g^*(t')dt' = \dot{\Psi}_0(\tau-t).
\label{eq:eleq}
\ee
In particular, the optimal irreversible work will be \cite{naze2022optimal}
\be
W_{\rm irr}^* = \frac{\delta\lambda^2}{2}\Psi_0(0)+\frac{\delta\lambda^2}{2}\int_0^\tau \dot{\Psi}_0(\tau-t)g^*(t)dt.
\ee

The objective of this work is to derive the linear response version of the variance of the work
\be
\sigma_{W}^2 = \langle \overline{W^2}\rangle_0-\langle \overline{W}\rangle_0^2
\ee
and to find an Euler-Lagrange equation of such a functional. In this way, the Euler-Lagrange equation will furnish the optimal protocol that will minimize the variance of the work. I shall accomplish it by using an extension to finite-time and weak processes of the fluctuation-dissipation relation \eqref{eq:wfdt}.

\section{Fluctuation-dissipation relation} 

To calculate the linear response version of the variance of the work, I remark that such quantity must be of second-order in the driving strength $\delta\lambda$. This occurs since the work starts in the first order and must be squared to calculate its variation. Therefore, it can be approximated as (see Appendix for details)
\begin{align}
\la \overline{W^2}\ra_0 &\approx \int_0^\tau\int_0^\tau \la \pd_\lambda\mc{H}(0)\overline{\pd_\lambda\mc{H}}(t-t')\ra_0\dot{\lambda}(t)\dot{\lambda}(t')dtdt',
\end{align}
To calculate $\la \overline{W}\ra_0^2$, the same reasoning is followed
\begin{align}
\la \overline{W}\ra_0^2 &\approx  \la \pd_\lambda\mc{H}(0)\ra_0^2.
\end{align}
Therefore, by rewriting the variance of the work in terms of the relaxation function, we have
\be
\sigma_W^2 \approx \frac{1}{\beta}\int_0^\tau\int_0^\tau \Psi_0(t-t')\dot{\lambda}(t)\dot{\lambda}(t')dtdt'+\mathcal{C}-\la \pd_\lambda\mc{H}(0)\ra_0^2
\label{eq:varLR}
\ee
The first term on the right-hand side is $2W_{\rm irr}/\beta$, while the second one, because the relaxation function decorrelates for long times, is (see Appendix for details)
\begin{align}
\mathcal{C}-\la \pd_\lambda\mc{H}(0)\ra_0^2 &= 0,
\end{align}
This immediately leads to the extended version of the work fluctuation-dissipation theorem 
\be
\overline{W}=\Delta F+\frac{\beta}{2}\sigma_W^2,
\label{eq:fdtheo}
\ee
which now holds for finite-time and weak processes. Observe that such a relation indicates that the work ceases to fluctuate for large switching times, as demonstrated in Ref.~\cite{yi2021}. Similar expressions for the first and second moments were also discussed in this article.

In the following I present an example that treats a case outside the hypotheses of Jarzynski's and Seifert's works \cite{jarzynski1997nonequilibrium,speck2004distribution}: a stochastic model with a time-dependent non-quadratic potential subjected to fast processes.

\section{Example}
\label{sec:example}

I am going to illustrate the relation \eqref{eq:wfdt} for a white noise overdamped Brownian particle, subjected to an anharmonic stiffening trap. First, I consider an overdamped Brownian particle in contact with a heat bath, whose dynamics are governed by the following Langevin equation
\be
\dot{x}(t)+\frac{1}{\gamma}\pd_x\mc{V}(x(t),\lambda(t)) =\eta(t),
\label{eq:langevin}
\ee  
where $x(t)$ is its position at the instant $t$, $\gamma$ is the damping coefficient, $\lambda(t)$ is the control parameter and $\eta(t)$ is a Gaussian white noise characterized by
\be
\overline{\eta(t)}=0, \quad \overline{\eta(t)\eta(t')}=\frac{2}{\gamma\beta}\delta(t-t').
\label{eq:bceta}
\ee
The time-dependent potential will be a quartic anharmonic stiffening trap
\be
\mc{V}(x(t),\lambda(t))=\frac{\lambda(t)}{4}x^4(t).
\label{eq:MovingLaserTrap}
\ee

\begin{figure}[t]
    \centering
    \includegraphics[scale=0.45]{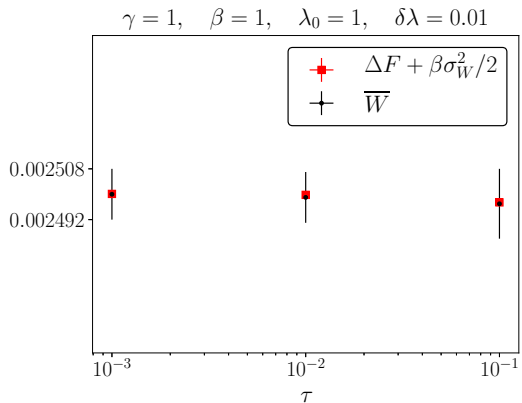}
    \caption{Verification of the work fluctuation-dissipation theorem of Eq. \eqref{eq:wfdt} for processes with $\tau=10^{-1},10^{-2},10^{-3}$. The black dots correspond to the values of $\overline{W}$ and the red squares to the values of $\Delta F+\beta\sigma_W^2/2$. We considered $\gamma=1$, $\beta=1$, $\lambda_0=1$ and $\delta\lambda=0.01$. The agreement is satisfactory.}
    \label{fig:2}
\end{figure}

To verify the work fluctuation-dissipation theorem, we numerically compare the average work $\overline{W}$ with the sum of $\Delta F$ and $\beta\sigma_W^2/2$. To do so, we consider a linear protocol, given by
\be
g(t) = \frac{t}{\tau},
\label{eq:linearprot}
\ee
and use the value of $\Delta F$ furnished by its analytical expression
\be
\Delta F = \frac{1}{4\beta}\ln{\left(1+\frac{\delta\lambda}{\lambda_0}\right)}.
\ee
Repeating $N_{\rm exp}$ experiments of sampling $N$ initial conditions in the canonical ensemble of the system for a stochastic process with a number of divisions $N_d$ of the switching time $\tau$, I collect the data for the analysis \cite{tome2015}. To achieve convergence in at least three significant digits in the relation \eqref{eq:wfdt}, we use $N_{\rm exp}= 10$, $N=2\times 10^5$ and $N_d=200$, for processes where $\tau=10^{-1},10^{-2},10^{-3}$. It was also used $\lambda_0=1$, $\delta\lambda=0.01$, $\gamma=1$ and $\beta=1$. It was assumed that in the approximation of 1\% in the perturbation linear-response theory would be in its range of validity, according to previous works~\cite{acconcia2015,naze2022kibble,kamizaki2022}. The results of plotting the averages of $\overline{W}$ and $\Delta F+\beta\sigma_W^2/2$ with respective standard deviations are depicted in Fig.~\ref{fig:2}. The agreement is satisfactory. The independence of the irreversible work on $\tau$ is indicative that the system is performing a fast process since no calculation of relaxation function was indeed made and a comparison with the relaxation time was not done~\cite{naze2022optimal}.

\section{Non-Gaussian probabilistic distribution}
\label{sec:nongaussian}

\begin{figure}[t]
    \centering
    \includegraphics[scale=0.45]{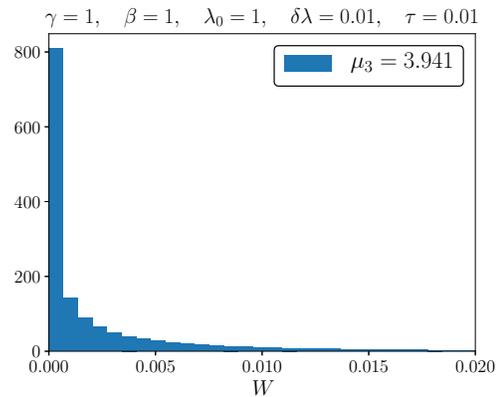}
    \caption{Non-Gaussian work probabilistic distribution for a process with $\tau=0.01, \lambda_0=1, \delta\lambda=0.01,\gamma=1,\beta=1$. The positive value of the skewness $\mu_3$ indicates the long right tail of the graphic.}
    \label{fig:3}
\end{figure}

By contrast with Ref.~\cite{jarzynski1997nonequilibrium,speck2004distribution}, the linear-response approach used in this approximation of the work probabilistic distribution is not a reduction to a Gaussian one. Indeed, Fig.~\ref{fig:3} depicts the work probabilistic distribution for a process with $\tau=0.01, \lambda_0=1, \delta\lambda=0.01,\gamma=1,\beta=1$. For this particular case, we have a non-Gaussian distribution, where its skewness is $\mu_3=3.941$, which indicates a long right tail and non-Gaussianity.

\section{Optimal work fluctuations} 
\label{sec:optimization}

Ever since it holds the fluctuation-dissipation relation for finite-time and weak processes, the functional of the variance of the work is the same (up to a factor $\beta/2$) as the one of the irreversible work. Therefore, the Euler-Lagrange equation that describes its optimization is the same as Eq.~\eqref{eq:eleq}. This means that, for finite-time and weak processes, if one achieves the minimal work spent, one does it in the precisest way possible. 

Consider as an example the white noise overdamped Brownian motion subjected to a harmonic stiffening trap. By looking at the optimal protocol that minimizes the irreversible work in Ref.~\cite{naze2022optimal}, the optimal protocol that minimizes the variance of work will be 
\be
g^*(t)=\frac{t+\tau_R}{\tau+2\tau_R},
\ee
with jumps at the beginning and end of the process, where $\tau_R$ is its relaxation time. The optimal work variance will be
\be
\sigma_W^{2*} = \frac{1}{\beta^2}\frac{{(\delta\lambda/\lambda_0)}^2}{\tau+2\tau_R}.
\ee
For this case, the variance is increased for fast processes, $\tau/\tau_R \ll 1$, or high temperatures, $\beta\ll 1$. As predicted by~\cite{yi2021}, the variance vanishes for slowly-varying processes, $\tau/\tau_R\gg 1$.

\section{Final remarks}
\label{sec:finalremarks}

In this work, using linear-response theory, I have extended the fluctuation-dissipation relation \eqref{eq:wfdt} for finite-time and weak processes. To illustrate such a result, I analyzed the example of the white noise overdamped Brownian motion subjected to an anharmonic stiffening trap. Such an extension shows that the optimal protocol that minimizes the irreversible work is the same that minimizes the variance of the work. By contrast with previous results in the literature for classical systems, the linear-response approach is not a Gaussian reduction in the work probabilistic distribution. This work opens the door for fluctuation-dissipation relations in higher orders of nonlinear response.

\bibliography{VARLR}

\onecolumngrid

\appendix
\section{Fluctuation-dissipation relation}

I derive first the second-order of $\la \overline{W^2}\ra $ in terms of $\delta\lambda$. One has
\begin{align}
\la \overline{W^2}\ra_0 &= \la \overline{\left(\int_0^\tau \pd_\lambda\mc{H}(t)\dot{\lambda}(t)dt\right)^2}\ra_0\\
&=\la \overline{\int_0^\tau\int_0^\tau \pd_\lambda\mc{H}(t_1)\pd_\lambda\mc{H}(t_2)\dot{\lambda}(t_1)\dot{\lambda}(t_2)dt_1dt_2}\ra_0\\
&= \int_0^\tau\int_0^\tau \la\overline{\pd_\lambda\mc{H}(t_1)\pd_\lambda\mc{H}(t_2)}\ra_0\dot{\lambda}(t_1)\dot{\lambda}(t_2)dt_1dt_2\\
&= \int_0^\tau\int_0^\tau \la \overline{e^{\mc{L}t_1}\pd_\lambda\mc{H}(0)e^{\mc{L}t_2}\pd_\lambda\mc{H}(0)}\ra_0\dot{\lambda}(t_1)\dot{\lambda}(t_2)dt_1dt_2\\
&\approx \int_0^\tau\int_0^\tau \la \overline{e^{\mc{L}_0 t_1}\pd_\lambda\mc{H}(0)e^{\mc{L}_0 t_2}\pd_\lambda\mc{H}(0)}\ra_0\dot{\lambda}(t_1)\dot{\lambda}(t_2),
\end{align}
where, in the time-evolution operator $e^{\mc{L}t}$ of the perturbed solution, I used its non-perturbed solution version
\be
e^{\mc{L}t} \approx e^{\mc{L}_0 t}
\ee
to furnish the second-order expansion of $\langle \overline{W^2}\rangle_0$. Therefore
\be
\la \overline{W^2}\ra_0\approx \int_0^\tau\int_0^\tau \la\overline{ \pd_\lambda\mc{H}(0)e^{\mc{L}_0(t_1-t_2)}\pd_\lambda\mc{H}(0)}\ra_0\dot{\lambda}(t_1)\dot{\lambda}(t_2)dt_1dt_2.
\ee

Using the same argument, one has for $\la \overline{W}\ra_0^2$
\begin{align}
\la \overline{W}\ra_0^2 &= \left(\int_0^\tau \la \overline{\pd_\lambda\mc{H}(t)}\ra_0\dot{\lambda}(t)dt\right)^2\\
&= \int_0^\tau\int_0^\tau \la \overline{\pd_\lambda\mc{H}(t_1)}\ra_0\la\overline{\pd_\lambda\mc{H}(t_2)}\ra_0\dot{\lambda}(t_1)\dot{\lambda}(t_2)dt_1dt_2\\
&\approx \int_0^\tau\int_0^\tau \la\pd_\lambda\mc{H}(0)\ra_0\la \pd_\lambda\mc{H}(0)\ra_0\dot{\lambda}(t_1)\dot{\lambda}(t_2)dt_1dt_2\\
&= \int_0^\tau\int_0^\tau \la \pd_\lambda\mc{H}(0)\ra_0^2\dot{\lambda}(t_1)\dot{\lambda}(t_2)dt_1dt_2\\
&=\la \pd_\lambda\mc{H}(0)\ra_0^2.
\end{align}

Finally, the difference $\mathcal{C}-\la \pd_\lambda\mc{H}(0)\ra_0^2$ is given by
\begin{align}
\mathcal{C}-\la \pd_\lambda\mc{H}(0)\ra_0^2 &= \la\pd_\lambda\mc{H}(0)\pd_\lambda\mc{H}(\infty)\ra_0-\la \pd_\lambda\mc{H}(0)\ra_0^2\\
&=\la\pd_\lambda\mc{H}(0)\ra_0\la\pd_\lambda\mc{H}(\infty)\ra_0-\la \pd_\lambda\mc{H}(0)\ra_0^2\\
&=\la \pd_\lambda\mc{H}(0)\ra_0^2-\la \pd_\lambda\mc{H}(0)\ra_0^2\\
&=0.
\end{align}

\end{document}